\documentstyle[preprint,prc,aps]{revtex}
\begin{document}
\title{Microscopic Model of the Time Like
Electromagnetic Form Factor of the Nucleon
\protect \footnote{Work supported by GSI Darmstadt and BMFT}}
\author{H. C. D\"{o}nges
\protect \footnote{Part of the dissertation of H. C. D\"{o}nges},
M. Sch\"{a}fer and U. Mosel \\
Institut f\"{u}r Theoretische Physik, Universit\"{a}t Giessen \\
	D-35392 Giessen, Germany}
\maketitle
\begin{abstract}
A microscopic model of the electromagnetic form factor of the nucleon is
developed in a hadronic framework, including pions, nucleons and the
$\Delta$-resonance explicitly. The space like on-shell form factors are
reproduced and predictions for the half off-shell dependence are made. The
impact of this off-shell dependence in the time like sector ($q^2<1$ GeV$^2$,
thus including the region of vector meson dominance) is of main interest in
this investigation.
\end{abstract}
\newpage
\section{Introduction}
Electromagnetic form factors of hadrons contain important information about
the intrinsic structure of these particles and about their coupling to the
external electromagnetic field. This electromagnetic structure has been
explored mainly by electron scattering experiments, giving information on the
electromagnetic form factor in the space like momentum region and thus on the
spatial distribution of charges and magnetic moments inside the hadrons in
general, and the nucleon in particular. The available data cover the momentum
range from $q^2 = 0$ up to about $q^2 = 35$ GeV$^2$ and can be well
parametrized by a dipole fit \cite{dipfit}. There exist models who explain the
data by assuming a certain number of poles which all, unfortunately, are in
the so-called "unphysical" region (not accessible by experiments on on-shell
nucleons) \cite{Hoehler2,Dub}.

The data are much more sparse for time like momentum transfers, where the
excitation of the nucleon and its decay is studied. Here the only available
data come from experiments at LEAR, exploiting $p\bar{p}$ annihilation
\cite{Bardin}. These data thus naturally start at momentum transfers larger
than two times the mass of the proton. They show that conventional pole fits
e.\ g.\ \cite{Dub} cannot be applied in this region any more. It is thus of
special interest, to explore the "unphysical region" and see whether there
exists a rich structure of poles and thus whether the so-called Vector Meson
Dominance (VMD \cite{Sakurai}), which is underlying these pole fits, is a
universal property of all hadrons or if it holds only for pions.

The region around momentum transfers corresponding to the vector meson mass
(around 750 - 800 MeV for $\rho$ and $\omega$ mesons) is so interesting
because these pole fits predict here a very pronounced resonance structure in
the form factors \cite{Dub}. Since access to this region for on-shell protons
is impossible, the only alternative is to look for half off-shell processes.
Indeed, dilepton production in hadronic collisions (bremsstrahlungs-dileptons)
offers access to the half off-shell electromagnetic form factor in the time
like region \cite{Schaefer1}.

Experiments of this kind have been performed by the DLS collaboration for
$p+p$, $p+A$ and $A+A$ collisions \cite{Roche}. Simulations of these processes
show for $p+p$ and $p+A$ a clear window for these bremsstrahlungs
contributions where these are not overshadowed by other processes \cite{Wolf};
for $A+A$ a strong $\omega$ peak is predicted \cite{Wolfpriv}.

Under the assumption of VMD the elementary dilepton production processes $NN
\to NN e^+ e^-$, $\pi N \to N e^+ e^-$ and $\gamma N \to N e^+ e^-$ were
studied by the Giessen group \cite{Schaefer1,Feuster,Schaefer2}. It is,
however, unclear if the time like electromagnetic form factor is influenced by
the off-shellness of the intermediate nucleon. The purpose of this paper is to
model the relevant vertex, to study its off-shell dependence and to
investigate if VMD is still visible if this off-shell dependence is properly
taken into account. The calculations are thus meant to stimulate and to
provide some guidance for experimental investigations of this important
hadronic property.

To obtain the general vertex, one needs a dynamical model that describes the
electromagnetic structure of the nucleon. It would be most desirable to use
the quark degrees of freedom. Unfortunately, state of the art quark models of
the nucleon do not allow to study off-shell effects and excitations
quantitatively, but only in a qualitative way. Their success is mainly
confined to space like properties of real nucleons at the present time
\cite{GCG}. In this paper, therefore, all calculations are performed using
hadronic degrees of freedom, following the concept of Naus and Koch \cite{NK}
or
Tiemeijer and Tjon \cite{TT}, who performed similar calculations for the space
like regime.

{}From a spectral analysis of form factors \cite{Hoehler} one has learned about
two important ingredients: $\pi N$-scattering and $\pi \pi$-scattering. While
$\pi\pi$-scattering is resonant, and thus in this channel best described by
taking into account a coupling to the $\rho$-meson, $\pi N$-scattering is more
difficult to implement, and will be handled only in a very schematic way in
the present model. On the other hand, semiphenomenological models
\cite{IJL,BRW} show the success of a description of the form factors in a
cloud/core picture. This will be discussed later in this paper.

Constrained by gauge invariance of the electromagnetic interaction, the vertex
is constructed from a pion loop expansion of the nucleon propagator by
coupling external photons to each charged particle in the loop. In this loop
expansion the $\Delta$-resonance is taken into account because, first of all,
it is necessary for the reproduction of $\pi N$-scattering, and secondly, it is
the first important resonance to contribute to the off-shell effects. In
addition to that, from an analysis of the Gerasimov-Drell-Hearn sum rule
\cite{Burkert}, the $\Delta$ is found to be most important to understand the
anomalous magnetic moments of the nucleons.

This paper is organized as follows. In section \ref{sec2}, the general
structure of form factors for nucleons and pions is discussed, gauge
invariance and its consequences are introduced. In section \ref{sec3}, the
model is presented, and gauge invariance is proven for a simple case. A
discussion of VMD in this context is given in section \ref{sec4}. Results and
comparison to data are shown in section \ref{sec5}. Section \ref{sec6}
concludes the paper.

\section{General Structure of Form Factors}
\label{sec2}
\subsection{Form Factor of Nucleons}
The most general form of the electromagnetic interaction vertex for nucleons
can be split into an isoscalar part and an isovector part
\begin{equation}
\Gamma^\mu (p^\prime,p) = \Gamma^\mu_S (p^\prime,p) + \tau_3 \Gamma^\mu_V
(p^\prime,p) \quad.
\end{equation}
The vertices of proton or neutron are then linear combinations
$\Gamma^\mu_p (p^\prime,p)=\Gamma^\mu_S (p^\prime,p)+\Gamma^\mu_V
(p^\prime,p)$,
$\Gamma^\mu_n (p^\prime,p)=\Gamma^\mu_S (p^\prime,p)-\Gamma^\mu_V
(p^\prime,p)$.

The isoscalar or isovector parts can be split up further. Let $p$ denote the
4-momentum of the incoming nucleon, $p^\prime$ the 4-momentum of the outgoing
nucleon, and $q=p-p^\prime$ the 4-momentum of the outgoing photon. This choice
is more convenient for the case of time like momentum transfer and for
dilepton production, but is different to what is conventionally used in the
literature, where $q$ is the 4-momentum of the incoming photon; this amounts
to some sign changes in the decomposition of the vertex compared to other
papers \cite{NK,TT,Bincer}. According to \cite{Bincer} the isoscalar/isovector
vertex can be expressed as
\begin{equation}
\Gamma^\mu_{S,V} (p^\prime,p)=e \sum_{r^\prime,r=0}^{1} \{ \gamma\cdot p^\prime
\}^{r^\prime}
\left(
        A^{r^\prime r}_{1 S,V} \gamma^\mu
       -A^{r^\prime r}_{2 S,V} \frac{\mbox{i} \sigma^{\mu \nu} q_\nu}{2 m}
       -A^{r^\prime r}_{3 S,V} q^\mu
\right) \{ \gamma\cdot p \}^r \quad.
\label{eq2_1_1}
\end{equation}
The 24 functions $A^{r^\prime r}_{i S,V} (i=1..3; r,r^\prime=0,1)$ are scalar
functions of the three variables $p^2$, ${p^\prime}^2$ and $q^2$.
Introducing a shorthand notation
\begin{displaymath}
{\cal O}^\mu_1 = \gamma^\mu ,\quad
{\cal O}^\mu_2 = -\frac{\mbox{i} \sigma^{\mu \nu} q_\nu}{2 m}   ,\quad
{\cal O}^\mu_3 = -q^\mu     ,\quad
\mbox{($m$ = nucleon mass)}
\end{displaymath}
and the projection operator
\begin{equation}
\Lambda^s(p) = \frac{s \gamma\cdot p + W}{2 W} \quad\mbox{with }
W = \sqrt{p^2} ,\quad
s = \pm 1 \quad,
\end{equation}
(\ref{eq2_1_1}) can be written as (drop $S$ and $V$ for convenience)
\begin{equation}
\Gamma^\mu (p^\prime,p) = e \sum_{i=1}^3 \sum_{s,s^\prime=\pm 1}
\Lambda^{s^\prime}(p^\prime) F^{s^\prime s}_{i} (W^\prime,W;q^2) {\cal O}^\mu_i
\Lambda^s(p) \quad.
\label{eq2_2_3}
\end{equation}
The functions $F^{s^\prime s}_i$ are linear combinations of the $A^{r^\prime
r}_i$.

Since for $p^2=m^2$ the operator $\Lambda^s(p)$ is just the projection operator
to
positive ($s=+1$) or negative ($s=-1$) energies, $\Gamma^\mu (p^\prime,p)$
taken
between
on-shell spinors reduces to
\begin{displaymath}
\bar{u}(p^\prime) \Gamma^\mu (p^\prime,p) u(p) =
e \sum_{i=1}^3 \bar{u}(p^\prime) F^{++}_i(m,m;q^2) {\cal O}^\mu_i u(p) \quad.
\end{displaymath}
It turns out that $F^{++}_3 (m,m;q^2) = 0$ (time reversal invariance), so
the on-shell vertex takes on the well known form \cite{BD}
\begin{displaymath}
\bar{u}(p^\prime) \Gamma^\mu (p^\prime,p) u(p) =
e \bar{u}(p^\prime) \left[ F^{++}_1(q^2) \gamma^\mu - F^{++}_2(q^2)
\frac{\mbox{i} \sigma^{\mu \nu} q_\nu}{2 m} \right] u(p) \quad.
\end{displaymath}
At $q^2=0$ the form factor $F^{++}_1$ measures the electric charge and
$F^{++}_2$ measures the anomalous magnetic moment.

The vertex $\Gamma^\mu (p^\prime,p)$ contains self energy corrections in the
external legs.
For this reason it is often called {\em reducible} vertex. In this paper it is
also called {\em full} vertex in order to indicate that in experimental
measurements these self energy corrections are always included. To eliminate
these corrections, the {\em irreducible} vertex is defined by
\begin{equation}
S(p^\prime) \Gamma^\mu_{irr} (p^\prime,p) S(p) = S_0(p^\prime) \Gamma^\mu
(p^\prime,p) S_0(p)  \quad.
\label{eq2_2_4}
\end{equation}
For the irreducible vertex a decomposition exists similar to the one for the
full vertex, as will be proven now. In (\ref{eq2_2_4}) $S_0(p)$ is the free
propagator, $S(p)$ is the full propagator, including all self energy
corrections $\Sigma(p) = S^{-1}_0(p) - S^{-1}(p)$. Most generally $\Sigma(p)$
is decomposed into a vector part $\Sigma_V(W)$ and a scalar part $\Sigma_S(W)$
by $\Sigma(p) = \Sigma_V(W) \gamma\cdot p - \Sigma_S(W) m$, so the full
propagator
takes on the form
\begin{displaymath}
  S^{-1}(p)=\gamma\cdot p-m-\Sigma(p)
= \gamma\cdot p (1-\Sigma_V(W)) - m (1-\Sigma_S(W))
= \gamma\cdot p^* - m^* \quad.
\end{displaymath}
It proves helpful to introduce the positive and negative energy projections of
the self energy and the propagator $\Sigma^\pm(W)$ and $S^\pm(W)$ by
\begin{displaymath}
\Sigma(p) = \sum_{s=\pm 1} \Lambda^s(p) \Sigma^s(W) ,\quad
 S(p)     = \sum_{s=\pm 1} \Lambda^s(p) S^s(W) \quad.
\end{displaymath}
Using the properties of the projection operators
\begin{equation}
\Lambda^+(p) + \Lambda^-(p) = 1 ,\quad
\Lambda^+(p) - \Lambda^-(p) = \frac{\gamma\cdot p}{W} \quad,
\end{equation}
one finds for the self energy
\begin{equation}
\Sigma^s(W) = s \Sigma_V(W) W - \Sigma_S(W) m
\end{equation}
and for the full propagator
\begin{equation}
S^s(W) = \frac{1}{s W^* - m^*} = \frac{1}{s W - m - \Sigma^s(W)} \quad,
\end{equation}
where $W^* = W ( 1-\Sigma_V(W) )$ and $m^* = m ( 1-\Sigma_S(W) )$. With this
notation one finds further
\begin{eqnarray}
         S_0(p)          = \sum_s \Lambda^s(p) S^s_0(W)
&;\quad& S^s_0(W)        = \frac{1}{sW-m}
\\
         S^{-1}(p)       = \sum_s \Lambda^s(p) {S^{-1}}^s(W)
&;\quad& {S^{-1}}^s(W)   = sW^*-m^*
\\
         S^{-1}_0(p)     = \sum_s \Lambda^s(p) {S^{-1}_0}^s(W)
&;\quad& {S^{-1}_0}^s(W) = sW-m \quad.
\end{eqnarray}
Now one can easily derive the relation between the reducible vertex
(\ref{eq2_2_3}) and the irreducible vertex (\ref{eq2_2_4}):
\begin{eqnarray}
\Gamma^\mu_{irr} (p^\prime,p) &=& S^{-1}(p^\prime) S_0(p^\prime) \Gamma^\mu
(p^\prime,p) S_0(p) S^{-1}(p) \nonumber \\
            &=& e \sum_{i=1}^3 \sum_{s^\prime s} \Lambda^{s^\prime}(p^\prime)
                \frac{s^\prime {W^\prime}^* - {m^\prime}^*}{s^\prime {W^\prime}
- m} F^{s^\prime s}_{i} (W^\prime,W;q^2)
                \frac{s W^* - m^*}{s W - m} {\cal O}^\mu_i \Lambda^s(p)
 \nonumber \\
            &=& e \sum_{i=1}^3 \sum_{s^\prime s} \Lambda^{s^\prime}(p^\prime)
f^{s^\prime s}_i (W^\prime,W;q^2) {\cal O}^\mu_i \Lambda^s(p) \quad.
\label{eq_gamirr}
\end{eqnarray}
Eq.\ (\ref{eq_gamirr}) represents the desired decomposition of the irreducible
vertex in terms of the irreducible form factors $f^{s^\prime s}_i
(W^\prime,W;q^2)$.

The Ward-Takahashi-identity (WTI) relates the irreducible vertex to the full
propagator \cite{WTI}:
\begin{equation}
q_\mu \Gamma^\mu_{irr} (p^\prime,p) = e \hat{Q} ( S^{-1}(p) - S^{-1}(p^\prime)
)
\quad.
\label{eq2_4_12}
\end{equation}
$\hat{Q}$ is the appropriate charge operator.

Two useful identities will be introduced here:
\begin{eqnarray}
\Lambda^{s^\prime}(p^\prime) \gamma\cdot q \Lambda^s(p) &=&
\Lambda^{s^\prime}(p^\prime) (s W - s^\prime {W^\prime}) \Lambda^s(p)
\label{eq2_4_13}\\
- \Lambda^{s^\prime}(p^\prime) \mbox{i} \sigma_{\mu \nu} q^\nu \Lambda^s(p)
&=& \Lambda^{s^\prime}(p^\prime) [(sW+s^\prime {W^\prime}) \gamma_\mu -
(p+p^\prime)_\mu] \Lambda^s(p) \quad. \label{eq2_4_14}
\end{eqnarray}
(\ref{eq2_4_14}) is a general form of the Gordon-identity \cite{BD}.
With the use of (\ref{eq2_4_13}) and $q_\mu {\cal O}^\mu_2 = 0$ the projection
of
(\ref{eq2_4_12}) onto positive and negative states yields
\begin{eqnarray}
f^{s^\prime s}_1 (sW-s^\prime {W^\prime}) - q^2 f^{s^\prime s}_3
&=& \hat{Q} \left( {S^{-1}}^s(W) - {S^{-1}}^{s^\prime}({W^\prime}) \right)
\nonumber \\
&=& \hat{Q} \left( (s W^* - s^\prime {W^\prime}^*) - (m^* - {m^\prime}^*)
\right) \quad.
\label{eq2_4_14a}
\end{eqnarray}
For the case of an outgoing on-shell particle (${W^\prime} = m$) of positive
energy
($s^\prime = +1$) this reduces to
\begin{equation}
f^{+s}_1 - \frac{q^2}{sW-m} f^{+s}_3 = \hat{Q} \frac{s W^* - m^*}{sW-m}
\label{eq2_4_15}
\end{equation}
or for the full form factor
{\Large(} (\ref{eq2_4_15}) $\cdot \frac{sW-m}{sW^*-m^*}$ {\Large)}:
\begin{equation}
F^{+s}_1 - \frac{q^2}{sW^*-m^*} F^{+s}_3 = \hat{Q} \quad.
\label{eq2_4_16}
\end{equation}
To get (\ref{eq2_4_16}) one needs further
$\left. \frac{{W^\prime}-m}{{W^\prime}^*-{m^\prime}^*}\right|_{{W^\prime}=m} =
1$, i.\ e.\ the full
propagator has a pole of unit residue at the physical mass $m$ of the nucleon.

For real photons ($q^2=0$) one recovers that $F^{+s}_1=1$ for protons and
$F^{+s}_1=0$ for neutrons. This holds for the full form factor even if the
incoming nucleon is off-shell. Note that (\ref{eq2_4_15}) implies that
$f^{+s}_1(q^2=0)$ depends on the nucleon's off-shellness and only reduces
to the real charge for on-shell incoming particles. The latter must be true
because for on-shell particles there is no difference between full and
irreducible vertex.

Note that the WTI does not pose any constraint on the magnetic form factors
$F^{+s}_2$. Note also, that $F^{+s}_3(m,m;q^2) = f^{+s}_3(m,m;q^2)$ have to
vanish for all $q^2$ in order to obtain finite contributions on the lhs of
(\ref{eq2_4_15}) and (\ref{eq2_4_16}).

To obtain the full vertex, the knowledge of the full propagator is needed,
which can be obtained from $f^{+s}_1$ using relation (\ref{eq2_4_15}) taken
at $q^2=0$:
\begin{displaymath}
{S^{-1}}^s(W) = sW^*-m^* = f^{+s}_{1,p}(m,W;q^2=0) (sW-m) \quad.
\end{displaymath}
The index $p$ stands for proton. This allows to write the half off-shell full
form factor as
\begin{equation}
F^{+s}_i(m,W;q^2) = \frac{f^{+s}_{i}(m,W;q^2)}{f^{+s}_{1,p}(m,W;0)} \quad.
\label{eq2_5_1}
\end{equation}
It is thus sufficient to calculate the irreducible vertex only.

{}From (\ref{eq2_4_14a}) more relations can be derived especially for $q^2=0$:
\begin{eqnarray*}
f^{++}_1(W,m;0) &=& f^{++}_1(m,W;0) \\
f^{+-}_1(m,W;0) &=& f^{-+}_1(W,m;0) \\
f^{--}_{1,p}(m,m;0) &=& 1 ,\quad f^{--}_{1,n}(m,m;0) = 0 \quad.
\end{eqnarray*}
So the off-shell full form factor is
\begin{equation}
F^{s^\prime s}_{i} (W^\prime,W;q^2) = \frac{f^{s^\prime s}_i
(W^\prime,W;q^2)}{f^{s^\prime+}_{1,p}({W^\prime},m;0) f^{+s}_{1,p}(m,W;0)}
\quad.
\label{eq2_6_1}
\end{equation}

Experiments always measure the full form factor. $F^{s^\prime s}_3$ is never
accessible by experiments since ${\cal O}^\mu_3 j_\mu = 0$ for any conserved
current
$j_\mu$.
%
%
\subsection{Form Factor of Pions}
\label{sec2_2}
The most general form of the pion photon vertex is \cite{GR}
\begin{displaymath}
\Gamma^\mu_\pi (p^\prime,p) = e \hat{Q}_\pi
[ f_1({p^\prime}^2,p^2;q^2) {p^\prime}^\mu + f_2({p^\prime}^2,p^2;q^2) p^\mu ]
\quad.
\end{displaymath}
A more convenient and also better known notation is
\begin{displaymath}
\Gamma^\mu_\pi (p^\prime,p) = e \hat{Q}_\pi
[ A({p^\prime}^2,p^2;q^2) P^\mu_L + B({p^\prime}^2,p^2;q^2) P^\mu_T ]
\end{displaymath}
with $P_L = p^\prime + p$, $P_T = P_L - q (P_L \cdot q / q^2)$, $q=p-p^\prime$.

If $j_\mu$ is a conserved current, one measures
\begin{displaymath}
j_\mu \Gamma^\mu_\pi (p^\prime,p) = e \hat{Q}_\pi
[ A({p^\prime}^2,p^2;q^2) + B({p^\prime}^2,p^2;q^2) ] j_\mu P^\mu_L \quad,
\end{displaymath}
i.\ e.\ the sum of $A$ and $B$. The WTI requires
\begin{displaymath}
q_\mu \Gamma^\mu_\pi (p^\prime,p) = e \hat{Q}_\pi q_\mu P^\mu_L
A({p^\prime}^2,p^2;q^2)
= e \hat{Q}_\pi [D^{-1}_\pi(p^2) - D^{-1}_\pi({p^\prime}^2)] \quad.
\end{displaymath}
Thus only $A$ is constrained by the WTI.

The on-shell form factor of pions is a measurable quantity. It is given to
good precision by the vector meson dominance (VMD) hypothesis \cite{Quenzer}.
In fact the VMD works so well that one is led to assume that the bare pion is
essentially a structureless particle \cite{Weise}. It is therefore safe to
neglect effects of the pion's off-shellness and to assume that $A$ and $B$
depend on $q^2$ only:
\begin{displaymath}
j_\mu \Gamma^\mu_\pi (p^\prime,p) = e \hat{Q}_\pi F_\pi(q^2) j_\mu P^\mu_L
\quad,
\end{displaymath}
where $F_\pi(q^2)$ is the measured form factor. This fixes $B(q^2)$:
\begin{displaymath}
B(q^2) = F_\pi(q^2) - A(q^2) \quad.
\end{displaymath}
Given any form of the pion self energy, this allows to maintain gauge
invariance as well as the measured form factor. Section \ref{sec4} will
explain the VMD hypothesis and will show how to carry it over to the nucleon.

\section{Model for the Form Factor}
\label{sec3}
As mentioned in the introduction, there exists a subtle interplay of $\pi N$-
and $\pi \pi$ scattering in describing the electromagnetic properties on
nucleons. This section is devoted to introduce the part of the model that is
suited to describe the $\pi N$ interaction. Subsection \ref{sec3_1} introduces
the Lagrangians for nucleons, pions and $\Delta$s. In a naive picture the
coupling of photons is introduced. Subsection \ref{sec3_2} describes the
coupling of the photon to these fields in the correct way and shows that this
is equivalent to the picture developed in subsection \ref{sec3_1}, which is
thus respecting the local U(1)-symmetry of QED. In section \ref{sec4} the
$\pi \pi$ interaction is modeled in terms of vector meson dominance.

\subsection{Interaction of Mesons, Baryons and Photons}
\label{sec3_1}
Part of the structure of the nucleon is due to the meson cloud that dresses
the bare nucleon. This idea is well established and several phenomenological
models exist that take it into account \cite{IJL,BRW}. For electromagnetic
form factors Naus et al.\ \cite{NK}, Tiemeijer et al.\ \cite{TT} or Bos et
al.\ \cite{Bos} performed detailed calculations based on a meson baryon
interaction picture, all giving essentially the same results for the shape and
the off-shell dependence of the form factors, but differing somewhat in their
predictions for the anomalous magnetic moments. However these authors
restricted themselves to space like momentum transfer and to off-shell nucleons
with $W<m+m_\pi$ in order to avoid poles due to decay into inelastic channels.
Since the purpose of this paper is to compute the form factor in the time like
region, the decay modes have to be included; also the $\Delta$-resonance will
be considered in the calculation because for off-shell nucleons with $W>m$ the
$\Delta$ is no longer kinematically supressed.

The model used here is based on the Lagrangian densitiy for pions and nucleons
with pseudovector coupling.
\begin{equation}
{\cal L}_N = \bar{\Psi} (\gamma\cdot p-m) \Psi
    + \frac12 [(\partial_\mu \tilde{\pi})^\dagger (\partial^\mu \tilde{\pi})
               -m_\pi^2 \tilde{\pi}^\dagger \tilde{\pi} ]
    + \frac{g_{NN\pi}}{2m} \bar{\Psi} \gamma_5 \gamma^\mu \tilde{\tau} \Psi
      \partial_\mu \tilde{\pi} \quad.
\label{eq3_1_1}
\end{equation}
Including the $\Delta$ leads to an additional term in
(\ref{eq3_1_1}):
\begin{equation}
{\cal L}_\Delta = \bar{\Psi}^\mu_\Delta \Lambda_{\mu \nu} \Psi^\nu_\Delta
           + \frac{g_{N\Delta\pi}}{2m} \left(
             \bar\Psi^\mu_\Delta \tilde{T} \Psi \partial_\mu \tilde{\pi}
             + \mbox{h.c.} \right)
\label{eq3_1_2}
\end{equation}
with
\begin{equation}
\Lambda_{\mu \nu} = (\gamma\cdot p - m_\Delta) g_{\mu \nu}
                  - (\gamma_\mu p_\nu + p_\mu \gamma_\nu)
                  + \gamma_\mu \gamma\cdot p \gamma_\nu
                  + m_\Delta \gamma_\mu \gamma_\nu \quad,
\end{equation}
as derived in the Rarita-Schwinger formalism \cite{Rarita,BW}, and $\tilde{T}$
being the matrix that couples isospin $3/2$ to isospin $1/2 \oplus 1$. The
$\Delta$ is here treated as a stable particle; its finite decay width must be
neglected at the order of diagrams discussed here. Using a momentum dependent
width would amount to including self energy corrections to the $\Delta$
propagator; this corresponds to diagrams of higher order in pion lines. Also a
whole set of new diagrams would be necessary to maintain gauge invariance. On
the other hand a constant width in the $\Delta$ propagator always yields
complex form factors, even for on-shell nucleons, because the relevant
thresholds are not taken into account.

This subsection will describe how the corrections to the electromagnetic
interaction vertex can be constructed for diagrams including one pion loop.
Starting point will be the nucleon propagator, which is up to the one pion
loop level given by fig.\ \ref{fig3_1}. In the naive picture the photon
couples to all charged particles individually. That this indeed fulfills the
WTI will be proven in the following subsection. The irreducible vertex is
given by fig.\ \ref{fig3_2}, where the coupling of the photon to the hadrons
is according to the usual Feynman rules. Since the pion coupling is chosen to
be pseudovector, additional contact terms arise (fig.\ \ref{fig3_2}d,e).

Since all the loop diagrams diverge, a regularization by a covariant cutoff of
the form
\begin{equation}
f_c(q^2) = \frac{m_\pi^2 - \Lambda^2}{q^2 - \Lambda^2} = (m_\pi^2 - \Lambda^2)
D_c(q)
\label{eq3_dc}
\end{equation}
is used. This form of the cutoff is chosen because it can be visualized as the
propagator $D_c(q)$ of a particle of mass $\Lambda$ with the same quantum
numbers as the pion; this ficticiuos {\bf c}ut{\bf o}ff-{\bf p}article will be
called "cop". The $NN\pi$-vertex is regulated with the monopole form, whereas
the $N\Delta\pi$-vertex needs a dipole cutoff to yield convergence because the
$\Delta$-propagator is ${\cal O}(p)$. In the spirit of the previous paragraph
the photon will then also couple to the cop if the intermediate pion carries
charge. The photon-pion coupling must therefore be replaced by a sum of
vertices given in fig.\ \ref{fig3_3}; the double dashed line denotes the
propagator of the cop. The diagrams with an internal $\Delta$-propagator have
two cop propagators because of the dipole cutoff. The pion-photon-vertex in
this case must thus be replaced by five diagrams.

The mass of the cop must be chosen large enough so that unphysical decay modes
are avoided. The possible decay modes and thus contributions to the imaginary
part of the form factor can be found by applying the usual cutting rules. If
the incoming nucleon is far enough off-shell ($W>m+m_\pi$) it can decay into
an on-shell nucleon and a real pion as depicted in fig.\ \ref{fig3_4}a. In
principle the off-shell nucleon can also decay into a nucleon and a cop if
$W>m+\Lambda$. This inequality sets the limit of the model. For the case of
this
paper $W_{max}$ will be restricted to 2 GeV, so $\Lambda$ should be somewhat
larger than 1 GeV. This is in good agreement with cutoff values used in meson
exchange potentials. If the cut is taken as indicated in fig.\ \ref{fig3_4}b
it becomes clear that there is also a restriction on the invariant mass $M$ of
the photon. If $M=\sqrt{q^2_\gamma}>2m_\pi$, $\pi^+\pi^-$-production becomes
possible. Pion-cop-production is not possible if $M$ is restricted to
$M<m_\pi+\Lambda \approx m_\pi+W_{max}-m \approx 1$ GeV.
%
%
\subsection{Ward-Takahashi-Identity}
\label{sec3_2}
This subsection is devoted to the proof of correctness of the picture
developed above. First, the appropriate electromagnetic interaction vertex
will be derived from the Lagrangians (\ref{eq3_1_1},\ref{eq3_1_2}) by minimal
coupling. Then the "reduced" formalism will be introduced to keep track of the
cop. Finally it will be proven that the WTI is fulfilled.

Since the photon is the gauge field of local $U(1)$ symmetry, it is introduced
into the Lagrangian by substituting $p_\mu \to p_\mu - e \hat{Q} A_\mu$, where
$\hat{Q}$ is the charge operator. This leads to the usual interaction terms for
nucleons and pions:
\begin{displaymath}
{\cal L}^N_{em}=-e \bar{\Psi} \gamma^\mu \hat{Q}_N \Psi A_\mu
         -\mbox{i} e \tilde{\pi}^\dagger \left(
          \stackrel{\rightarrow}{\frac{\partial}{\partial x_\mu}}
          -\stackrel{\leftarrow}{\frac{\partial}{\partial x_\mu}}
         \right) \hat{Q}_\pi \tilde{\pi} A_\mu \quad,
\end{displaymath}
to a contact term because of the derivative in the pseudovector coupling of
pions and nucleons:
\begin{displaymath}
{\cal L}^c_{em} = -e \frac{g_{NN\pi}}{2m}
              \bar{\Psi} \gamma_5 \gamma^\mu \tilde{\tau} \Psi
              \hat{Q}_\pi \tilde{\pi} A_\mu \quad,
\end{displaymath}
and to terms from ${\cal L}_\Delta$:
\begin{eqnarray*}
{\cal L}^\Delta_{em} &=& -e \bar{\Psi}^\mu_\Delta
                      (\gamma_\lambda g_{\mu \nu}
                      +\gamma_\mu g_{\lambda \nu}
                      +\gamma_\nu g_{\lambda \mu}
                      -\gamma_\mu \gamma_\lambda \gamma_\nu)
                      \hat{Q}_\Delta \Psi^\nu_\Delta A^\lambda
\\
                 &&  -e \frac{g_{N\Delta\pi}}{2m} \left(
                     \bar{\Psi}^\mu_\Delta \tilde{T} \Psi
                     \hat{Q}_\pi \tilde{\pi} A_\mu + \mbox{h.c.} \right) \quad.
\end{eqnarray*}
The operators $\hat{Q}_{\pi,N,\Delta}$ return the appropriate charge. Since the
$\Delta$ can decay into a $\gamma$ and a nucleon, an additional term must be
included in the Lagrangian:
\begin{displaymath}
{\cal L}_{\Delta \to n \gamma} = \mbox{i} \frac{g_{\Delta N\gamma}}{2 m}
( \bar{\Psi}^\mu_\Delta \gamma_5 \gamma^\nu T^3 \Psi F_{\mu \nu}
+ \mbox{h.c.} ) \quad.
\end{displaymath}
The corresponding vertex is
$\gamma_5 (\gamma^\nu q^\mu - \gamma\cdot q g^{\mu \nu}) G_1$ where the factor
$\sqrt{2/3}$ from $T^3$ and the coupling constant have been absorbed in $G_1$.
According to \cite{Scadron} this vertex is mainly responsible for the M1
multipole which dominates the decay. $q$ is the momentum of the outgoing
photon. Note that the index $\mu$ of the above vertex contracts with the
$\Delta$, while the index $\nu$ contracts with the photon field.

To prove gauge invariance of the model developed in this paper, three steps
need to be done. To obtain a gauge invariant coupling to the pion, the
coupling to the cop needs to be investigated, as well. So, first, the
cop-photon interaction vertex is constructed, then, as a second step, the
effective pion-photon vertex is constructed to fulfill the WTI locally using
the "reduced" formalism of Gross and Riska \cite{GR}. The third step is to
actually prove the WTI for the photon-nucleon interaction.

Since the cop is not a fundamental particle, it does not appear in the
Lagrangian. Therefore, so far it does not interact with the electromagnetic
field. To obtain gauge invariance it needs the same interaction vertex as the
pion:
\begin{displaymath}
\Gamma^\mu_{cc\gamma} (p^\prime,p) = \Gamma^\mu_{\pi\pi\gamma} (p^\prime,p) = e
\hat{Q}_\pi (p+p^\prime)^\mu \quad,
\end{displaymath}
where $p$ and $p^\prime$ are the momenta of the incoming and outgoing particles
respectively. Note that the so constructed vertex obviously fulfills the
WTI of a free particle:
\begin{equation}
(p-p^\prime)_\mu \Gamma^\mu_{cc\gamma} (p^\prime,p)
= e \hat{Q}_\pi (p^2-{p^\prime}^2)
= e \hat{Q}_\pi\left( D^{-1}_c(p)-D^{-1}_c(p^\prime) \right) \quad.
\label{eq3_wti}
\end{equation}

Next the "reduced" formalism is introduced. This part of the paper follows
closely \cite{GR}, who treat the cutoff function $f_c$ as a contribution to
the pion polarization function, which is possible because $f_c$ is a function
of the pion momentum only. The "reduced" pion propagator $D_R(p)$ is defined
as
\begin{displaymath}
D_R(p) = f_c(p^2) D_\pi(p) f_c(p^2) \equiv \frac{1}{p^2-m_\pi^2-\Pi(p^2)}
\quad,
\end{displaymath}
which defines the polarization function to be
\begin{displaymath}
\Pi(p^2) = (p^2-m_\pi^2) (1-f^{-2}_c(p^2)) \quad.
\end{displaymath}
It has the necessary properties $\Pi(p^2=m_\pi^2)=0$,
$\partial \Pi / \partial p^2|_{p^2=m_\pi^2}=1-f^{-2}_c(m_\pi^2)=0$ because
$f_c(m_\pi^2)=1$. Thus $D_R(p)$ can be viewed as the {\em full} propagator in
this model.

The WTI requires a relation between the {\em full} propagator and the
{\em irreducible} vertex:
\begin{displaymath}
(p-p^\prime)_\mu \Gamma^\mu_{irr} (p^\prime,p)
= e \hat{Q}_\pi \left( D^{-1}_R(p) - D^{-1}_R(p^\prime) \right) \quad.
\end{displaymath}
Multiplying with the full propagator on each side gives the requirement
\begin{equation}
(p-p^\prime)_\mu D_R(p^\prime) \Gamma^\mu_{irr} (p^\prime,p) D_R(p)
= e \hat{Q}_\pi \left( D_R(p^\prime) - D_R(p) \right) \quad.
\label{eq3_4_4}
\end{equation}
Requirement (\ref{eq3_4_4}) can be satisfied with the following choice for the
irreducible vertex:
\begin{eqnarray*}
\Gamma^\mu_{irr} (p^\prime,p) &=& (m_\pi^2-\Lambda^2)^2 \times  \\
&& \left[ \Gamma^\mu_{cc\gamma} (p^\prime,p) D_c(p) D_\pi(p)
+ D_\pi(p^\prime) \Gamma^\mu_{\pi\pi\gamma} (p^\prime,p) D_\pi(p)
+ D_\pi(p^\prime) D_c(p^\prime) \Gamma^\mu_{cc\gamma} (p^\prime,p) \right]
\quad,
\end{eqnarray*}
where $\Gamma^\mu_{cc\gamma} (p^\prime,p)$ and $\Gamma^\mu_{\pi\pi\gamma}
(p^\prime,p)$ are the individual
vertices of the free particles. This is pictorially represented in
fig.\ \ref{fig3_3} and can be shown as follows:
\begin{eqnarray*}
\lefteqn{(p-p^\prime)_\mu D_R(p^\prime) \Gamma^\mu_{irr} (p^\prime,p) D_R(p)}
&&
\\
&=& (m_\pi^2-\Lambda^2)^2 e \hat{Q}_\pi
\left[ D_c(p^\prime) (D^{-1}_c(p)-D^{-1}_c(p^\prime)) D_c(p) D_\pi(p) D_c(p)
\right.
\\
&& \phantom{(m_\pi^2-\Lambda^2)^2 e \hat{Q}_\pi}
+ D_c(p^\prime) D_\pi(p^\prime)(D^{-1}_\pi(p)-D^{-1}_\pi(p^\prime)) D_\pi(p)
D_c(p)
\\
&& \phantom{(m_\pi^2-\Lambda^2)^2 e \hat{Q}_\pi}
+ \left. D_c(p^\prime) D_\pi(p^\prime) D_c(p^\prime)
(D^{-1}_c(p)-D^{-1}_c(p^\prime)) D_c(p)
\right]
\\
&=& (m_\pi^2-\Lambda^2)^2 e \hat{Q}_\pi \left[
     D_c(p^\prime)  D_\pi(p^\prime)  D_c(p^\prime)
   - D_c(p)     D_\pi(p)     D_c(p) \right]
\\
&=& e \hat{Q}_\pi (D_R(p^\prime)-D_R(p)) \quad,
\end{eqnarray*}
using the definition of $D_c(p)$ in (\ref{eq3_dc}) and the fact that the free
vertex fulfills the WTI with the free propagator (\ref{eq3_wti}).

The naive picture, developed in the previous subsection, to couple a photon
to each of the charged particles then very naturally emerges from the
requirement of gauge invariance. This simplifies the numerical treatment of
such processes considerably because the self energy of the pion need not be
constructed explicitly; only free propagators occur, and, therefore, the pole
structure is much more transparent.

On the other hand, the "reduced" formalism provides a simple way to prove
gauge invariance of the model. This is demonstrated for the exemplary case of
a nucleon-pion-loop, where, in order to reduce the effort, the pseudoscalar
coupling is used. The results also hold for the pseudovector case if the
contact terms are taken into account. For the proof only two diagrams need to
be considered: the photon couples to the internal nucleon
(fig.\ \ref{fig3_2}b), and the photon couples to the internal pion
(fig.\ \ref{fig3_2}c). The pion propagator must be replaced by the reduced
propagator and the vertex correspondingly by the irreducible vertex. For the
moment the isospin factors are neglected, at the end they will be considered
for protons and neutrons separately. With $q_\mu=(p-p^\prime)_\mu$ on gets for
diagram \ref{fig3_2}c
\begin{eqnarray}
&&q_\mu \int \frac{d^4k}{(2\pi)^4} g \gamma_5 \mbox{i} S_0(k) g \gamma_5
               \mbox{i}  D_R(p^\prime-k)
             (-\mbox{i}) \Gamma^\mu_{irr}(p^\prime-k,p-k)
               \mbox{i}  D_R(p-k)
\nonumber \\
&&\mbox{using (\ref{eq3_4_4})} \nonumber \\
&=& e \int \frac{d^4k}{(2\pi)^4} g \gamma_5  \mbox{i} S_0(k) g \gamma_5
             \mbox{i} \hat{Q}_\pi (D_R(p^\prime-k) - D_R(p-k)) \quad.
\label{eq3_6_5}
\end{eqnarray}
(\ref{eq3_6_5}) is equal to the difference of self energy diagrams. But since
$\hat{Q}_\pi$ is still left in the expression, it is only that part of the self
energy which is due to charged pions. The remaining contribution of the
neutral pion comes from diagram \ref{fig3_2}b:
\begin{eqnarray}
&&q_\mu \int \frac{d^4k}{(2\pi)^4} g \gamma_5 \mbox{i} S_0(p^\prime-k)
                        (-\mbox{i} e \hat{Q}_N \gamma^\mu)
                          \mbox{i} S_0(p-k) g \gamma_5
                          \mbox{i}  D_R(k)
\nonumber \\
&=&-e \int \frac{d^4k}{(2\pi)^4} g \gamma_5 S_0(p^\prime-k)
                       \hat{Q}_N(S^{-1}_0(p-k) - S^{-1}_0(p^\prime-k))
                        S_0(p-k) g \gamma_5
                        D_R(k)
\nonumber \\
&=& e \int \frac{d^4k}{(2\pi)^4} g \gamma_5  \mbox{i}
\hat{Q}_N(S_0(p^\prime-k)-S_0(p-k)) g \gamma_5
                         \mbox{i} D_R(k) \quad.
\label{eq3_6_6}
\end{eqnarray}
For an incoming proton only the $\pi^+n$-loop contributes to (\ref{eq3_6_5})
and gives an isospin factor 2, the $\pi^0p$-loop contibutes to (\ref{eq3_6_6})
with an isospin factor 1. Adding both, one finds exactly
$e(\Sigma(p^\prime)-\Sigma(p))$. For an incoming neutron the $\pi^-p$-loop
contributes to both diagrams with a factor of 2, however, since the charge
operators are present, there is a relative minus sign between (\ref{eq3_6_5})
and (\ref{eq3_6_6}), such that they cancel exactly. The $\pi^0n$-loop
contributes to neither diagram.

The WTI requires for the nucleon
\begin{eqnarray*}
q_\mu \Gamma^\mu_{NN\gamma} (p^\prime,p) &=& e \hat{Q}_N
(S^{-1}(p)-S^{-1}(p^\prime)) \\
&=& e \hat{Q}_N (\gamma\cdot p - m - \Sigma(p)
             -\gamma\cdot p^\prime + m + \Sigma(p^\prime))                  \\
&=& e \hat{Q}_N (\gamma\cdot q + \Sigma(p^\prime) - \Sigma(p)) \quad.
\end{eqnarray*}
Obviously the direct term (fig.\ \ref{fig3_2}a) accounts for the $\gamma\cdot
q$
whereas the vertex corrections exactly make up for the self energy. This
proves gauge invariance up to the desired order in the strong coupling
constant. The proof for pseudovector coupling or for the case of an internal
$\Delta$ follows the same scheme as outlined above.

With the presented method the electromagnetic vertex correction can be
constructed from the self energy given to the desired order in the strong
coupling constant. It is given by just adding external photon lines to each
propagator corresponding to a charged particle. If a cutoff function is needed
it can be absorbed in the "reduced" formalism, which is powerful enough to
allow a proof of gauge invariance, on one hand, and simple to implement
numerically, on the other hand.

\section{Role of Vector Meson Dominance}
\label{sec4}
The idea of vector meson dominance was first introduced by Sakurai
\cite{Sakurai}. It supposes that the photon couples to the hadron by first
converting to a vector meson (a $\rho$-meson specifically), which then couples
to the hadron. The idea was first investigated for the electromagnetic form
factor of the pion, where it turned out to be successful. For time like
momentum transfers the structure of the $\rho$-meson propagator shows up
clearly. Furthermore, based on the assumption that the $\rho$-meson is the
"gauge-boson" of local isospin rotation, Sakurai introduced the vector meson
universality hypothesis, which states that the $\rho$-meson couples to all
hadrons with the same universal coupling constant $g_\rho$, which is equal to
the $\rho$-$\gamma$ coupling constant $g_{\rho \gamma}$. If this was true, the
electromagnetic form factor of the proton should have the same shape as the
one of the pion.

{}From electron scattering on nucleons one knows that the space like form
factors are well described by the dipole fit \cite{dipfit}. This is not in
agreement with the simple assumption of VMD, which always results in monopole
form factors. For this reason a number of resonances besides the $\rho$-meson
are introduced. Their coupling constants to the nucleon are determined by a
fit to the data \cite{Hoehler2,Dub}, in a so-called pole fit. In the
calculations of ref.\ \cite{Hoehler2}, furthermore, information of $\pi
N$-scattering is included. By constraining the coupling constants it is
possible to eliminate the leading monopole term and thus to obtain a dipole
shape

These pole fits can be analytically continued into the unphysical region
(nucleon on mass shell, time like momentum transfer of the photon below the
$N \bar{N}$ threshold). A rich, dominant pole structure shows up, which - if
it were observable - would contribute a direct proof of VMD for the nucleon.
The pole fits also predict the behaviour of the form factors in the time like
region above $N\bar{N}$-threshold. Recent data taken at LEAR \cite{Bardin}
show disagreement with the fit obtained by \cite{Dub} on the basis of the VMD
hypothesis.

It was pointed out by H\"{o}hler et al.\ \cite{Hoehler}, that besides the poles
in
$\pi^+ \pi^-$-annihilation, also the process $N\bar{N} \to \pi^+ \pi^-$ is
important to describe the form factor of the nucleon. The failiure of the pole
fit of \cite{Dub} most probably is linked to this. It is thus an interesting
question to investigate if $\pi N$-scattering influences the picture of VMD
considerably.

Since the information is hidden in the "unphysical region", the experiments
must be performed using off-shell nucleons. It must thus be checked, if the
off-shell dependence of the form factor interferes with the VMD picture. In
this section a scheme is set up to model VMD for pions as well as for
nucleons, which in section \ref{sec5} will be combined with the off-shell
information.

\subsection{Coupling of Photons to Pions}
The electromagnetic form factor of the pion can be parametrized as
\begin{equation}
F_\pi(q^2) = \frac{m_\rho^2}{m_\rho^2-\mbox{i} m_\rho \Gamma_\rho(q^2) -q^2}
\quad,
\label{eq4_1_1}
\end{equation}
where $m_\rho=0.77$ GeV is the mass of the $\rho$-meson, and
\begin{displaymath}
\Gamma_\rho(q^2) = 0.2458 \frac{(q^2-4 m_\pi^2)^{3/2}}{q^2} \Theta(q^2\ge
4m_\pi^2)
\end{displaymath}
describes the decay of the $\rho$-meson into two pions, which is with more
than 99 \% the dominant decay-channel.

The coupling that leads to such a form factor can be modeled by the following
Lagrangian density. Since the only processes of interest for this discussion
are those competing with the coupling of a photon to the hadrons, only the the
neutral vector mesons $\rho^0$ and $\omega$ will be considered.
\begin{eqnarray}
{\cal L} &=& \phantom{-}\frac12
\left[ \left(
\partial_\mu + \mbox{i} (g_{\rho \pi} \rho_\mu + e A_\mu) T_3 + g_{\omega \pi}
\omega_\mu
\right) \tilde{\pi} \right]^\dagger \times
\nonumber \\
&& \phantom{-}\phantom{\frac12}\left[ \left(
\partial^\mu + \mbox{i} (g_{\rho \pi} \rho^\mu + e A^\mu) T_3 + g_{\omega \pi}
\omega^\mu
\right) \tilde{\pi} \right]
\nonumber \\
&& -\frac12 m_\pi^2 \tilde{\pi}^\dagger \tilde{\pi}
   -\frac14 F_{\mu \nu} F^{\mu \nu}
   + {\cal L}_{\rho \gamma} + {\cal L}_{\omega \gamma}
\\
{\cal L}_{\rho \gamma} &=&
   -\frac14 G^\rho_{\mu \nu} G^{\mu \nu}_\rho
   + \frac12 m_\rho^2 \rho^\dagger_\mu \rho^\mu
   -\frac{e}{2 g_{\rho \gamma}} F_{\mu \nu} G^{\mu \nu}_\rho
\label{eq4_1_2}\\
{\cal L}_{\omega \gamma} &=&
   -\frac14 G^\omega_{\mu \nu} G^{\mu \nu}_\omega
   + \frac12 m_\omega^2 \omega^\dagger_\mu \omega^\mu
   -\frac{e}{2 g_{\omega \gamma}} F_{\mu \nu} G^{\mu \nu}_\omega \quad,
\label{eq4_1_3}
\end{eqnarray}
where $G^\rho_{\mu \nu} = \partial_\mu \rho_\nu - \partial_\nu \rho_\mu$,
$G^\omega_{\mu \nu} = \partial_\mu \omega_\nu - \partial_\nu \omega_\mu$,
$F_{\mu \nu} = \partial_\mu A_\nu - \partial_\nu A_\mu$, $g_{\rho \gamma}$ and
$g_{\rho \pi}$
are the coupling constants of the $\rho$-meson to the photon and the pion
respectively, $g_{\omega \gamma}$ and $g_{\omega \pi}$ for the $\omega$-meson
accordingly.

Note that with the tensor coupling in the Lagrangian (\ref{eq4_1_2},
\ref{eq4_1_3}) the photon-vector meson vertex turns out to be proportional to
$q^2$. In this case the contributions of the vector mesons to the photon
polarization function vanish for $q^2=0$; the photon thus remains massless
very naturally, whereas in the other case more involved constructions are
needed to keep the photon mass zero \cite{Sakurai}. Also the coupling of the
field tensors is gauge invariant by definition.

For the form factor one gets
\begin{equation}
F_\pi(q^2) = 1
           + \frac{g_{\rho \pi}}{g_{\rho \gamma}} \frac{q^2}{m_\rho^2-\mbox{i}
m_\rho \Gamma_\rho(q^2)-q^2}
           + \frac{g_{\omega \pi}}{g_{\omega \gamma}}
\frac{q^2}{m_\omega^2-\mbox{i} m_\omega \Gamma_\omega(q^2)-q^2} \quad.
\label{eq4_2_1}
\end{equation}
Complete vector meson dominance (VMD) assumes that $g_{\rho \pi}=g_{\rho
\gamma}=g_\rho$
\cite{Sakurai} and neglects the coupling to the $\omega$-meson completely.
Under this assumption (\ref{eq4_2_1}) reduces to (\ref{eq4_1_1}). The constants
$g_{\rho \pi}$ and $g_{\rho \gamma}$, however, can be inferred from the
decay-properties of the
$\rho$-meson $\rho \to \pi^+ \pi^-$ and $\rho \to \gamma \to \mbox{e}^+
\mbox{e}^-$. By converting the measured widths \cite{PDG} into coupling
constants one finds $g_{\rho \pi}$ = 5.9 and $g_{\rho \gamma}$ = 5.1. So the
universality criterion is almost fulfilled, but not exactly.

To calculate the ratio $g_{\omega \pi}/g_{\omega \gamma}$ one uses
\begin{displaymath}
  \frac{g_{\rho \pi}^2/g_{\rho \gamma}^2}{g_{\omega \pi}^2/g_{\omega \gamma}^2}
= \frac{\Gamma_{\rho \to \pi \pi} \cdot \Gamma_{\rho \to \mbox{e}^+
\mbox{e}^-}}
       {\Gamma_{\omega \to \pi \pi} \cdot \Gamma_{\omega \to \mbox{e}^+
\mbox{e}^-}}
= \frac{1}{0.017^2} \quad.
\end{displaymath}
This shows that the $\omega$-contribution to the form factor is much smaller
than that of the $\rho$. Nevertheless, it shows up in the form factor because
of the rather small $\omega$-width which results in the structure on top of
the wide bump stemming from the $\rho$-meson (see fig.\ \ref{fig4_1}). The
picture that emerges is thus a little different from complete VMD: best
agreement with experiment is reached if one relaxes the assumption of vector
meson universality and allows to couple all particles, the photon and the
vector mesons, to the hadrons with the appropriate coupling constants. In
principle also the phases could be chosen to correctly describe the
$\rho$-$\omega$ interference \cite{dipfit}. However, for the arguments in
section \ref{sec5} the current agreement of data and theory is sufficient.
%
%
\subsection{Coupling of Photons to Baryons}
\label{sec4_2}
In this subsection the previously developed method will be carried over to
baryons, and to the proton in particular. As stated in section \ref{sec2} the
electromagnetic form factor has an isoscalar and an isovector piece. By the
VMD hypothesis these are related to the isoscalar $\omega$-meson and to the
isovector $\rho$-meson respectively. If one assumes a Lagrangian density
\begin{eqnarray*}
{\cal L} &=&
\bar{\Psi} \left( \gamma \cdot(\mbox{i} \partial -g_{\rho N} \tau_3 \rho^0
-g_{\omega N} \omega -\frac12 (1+\tau_3) e A) - m \right) \Psi \\
&&+ {\cal L}_{\rho \gamma} + {\cal L}_{\omega \gamma} - \frac14 F^{\mu \nu}
F_{\mu \nu}
\end{eqnarray*}
with ${\cal L}_{\rho \gamma}$ and ${\cal L}_{\omega \gamma}$
from (\ref{eq4_1_2}, \ref{eq4_1_3}), one finds
\begin{eqnarray}
F^{s^\prime s}_{i,^{proton}_{neutron}} (W^\prime,W;q^2)
&=& \phantom{+}F^{s^\prime s}_{i,S} (W^\prime,W;q^2) \pm F^{s^\prime s}_{i,V}
(W^\prime,W;q^2) \nonumber \\
&& + F^{s^\prime s}_{i,\omega} (W^\prime,W;q^2)
\frac{q^2}{m_\omega^2-\mbox{i} m_\omega \Gamma_\omega(q^2) - q^2} \nonumber \\
&& \pm F^{s^\prime s}_{i,\rho} (W^\prime,W;q^2)
\frac{q^2}{m_\rho^2-\mbox{i} m_\rho \Gamma_\rho(q^2) - q^2} \quad.
\label{eq4_3_2}
\end{eqnarray}
$F^{s^\prime s}_{i,S}$ and $F^{s^\prime s}_{i,V}$ contain the direct coupling
to the photon, $F^{s^\prime s}_{i,\omega/\rho}$ contain the coupling to the
vector mesons with the appropriate coupling constants. The structure of this
Lagrangian density is guided by a gauge principle and thus minimal coupling
for the vector meson fields. Additionally there exists a tensor coupling of
the form
\begin{equation}
{\cal L}^{tensor} =
 g_{\rho N} \bar{\Psi} \kappa_\rho \frac{\sigma^{\mu \nu} G^\rho_{\mu \nu}}{4
m}
\Psi
+g_{\omega N} \bar{\Psi} \kappa_\omega \frac{\sigma^{\mu \nu} G^\omega_{\mu
\nu}}{4 m} \Psi
\quad.
\label{eq_p13}
\end{equation}

As a useful example the most simple case with the assumption of no further
substructure of the nucleon will now be discussed. It is defined by the
Lagrangian given above and neglects all couplings to further mesons like
pions. Section \ref{sec5} will combine the picture developed in section
\ref{sec3} with the ideas given below to the complete scenario.

For this simple example one finds $F^{s^\prime s}_{i,S} (W^\prime,W;q^2) =
F^{s^\prime s}_{i,V} (W^\prime,W;q^2) = 1/2$,
$F^{s^\prime s}_{i,\omega} (W^\prime,W;q^2)=g_{\omega N}/g_{\omega \gamma}$ and
$F^{s^\prime s}_{i,\rho} (W^\prime,W;q^2)=g_{\rho N}/g_{\rho \gamma}$. In
principle there will
be contributions to $F_2$ by the tensor-coupling of the
vector mesons, but
these contributions cannot account for the anomalous magnetic moments of
proton or neutron since they are weighted with $q^2$ and thus do not appear
for real photons. So the anomalous magnetic moments are genuinely due to the
inner structure of the nucleons.

The ratios of the coupling constants can be determined with the help of
meson-exchange-potentials for the nucleon-nucleon-interaction, if one takes
e.g. \cite{Gross} one finds $g_{\rho N}/g_{\omega N} \approx 0.26$. From the
ratios of the
decay widths one infers $g_{\rho \gamma}/g_{\omega \gamma} =
\sqrt{\Gamma_{\omega \to \mbox{e}^+\mbox{e}^-} / \Gamma_{\rho \to
\mbox{e}^+\mbox{e}^-}} \approx 0.3$. These
ratios are remarkably close to each other, and will in this crude
approximation assumed to be equal. Also from \cite{Gross} one finds
$g_{\rho N}/g_{\rho \pi} \approx 0.47$ which is surprisingly close to $1/2$. It
must be
emphasized, even though this is only a qualitative discussion, the obtained
numbers are close to what is obtained by other authors (table \ref{tbl2}).
So as a summary one has for the proton
\begin{displaymath}
F_{1,proton}(q^2) = 1
           + \frac12 \frac{q^2}{m_\omega^2-\mbox{i} m_\omega
\Gamma_\omega(q^2)-q^2}
           + \frac12 \frac{q^2}{m_\rho^2-\mbox{i} m_\rho \Gamma_\rho(q^2)-q^2}
\quad.
\end{displaymath}
For space like $q^2$ and under the further assumption
$m_\rho \approx m_\omega \approx m_V$ one finds
$F_{1,proton}(q^2)=m^2_V / (m^2_V - q^2)$, and $F_{1,neutron}(q^2)= 0$. This
establishes the commonly believed VMD hypothesis. The result falls short to
explain the well established dipole fit to the electromagnetic form factor
\cite{dipfit} since it is only of monopole structure.

Concluding, one can state that the pion is essentially structureless, and that
probably all of its size, if tested electromagnetically, is due to the
$\rho$-meson. The nucleons obviously have structure besides the one due to
VMD. This is very much in the spirit of Iachello, Jackson and Lande
\cite{IJL}, who introduce a further function to take into account the
shortrange part of the interaction, or of the two-phase model of Brown, Rho
and Weise \cite{BRW}, who explicitly introduce contributions of the
quark-core. Also a lot of work has been done to model the structure of the
nucleon by taking into account the meson cloud in terms of the exchange of
virtual pions \cite{NK,TT}.
%
%
\subsection{Gauge invariance and VMD}
At first sight VMD spoils gauge invariance, since even for the free vertex and
propagator the WTI is no longer fulfilled in the simple model developed
above:
\begin{displaymath}
q_\mu \Gamma^\mu = e \gamma\cdot q \frac{m^2_V}{m^2_V - q^2} \not= e
\gamma\cdot
q \quad.
\end{displaymath}
This can be restored by using the same technique as for the pion in section
\ref{sec2_2}. Instead of the $\gamma^\mu$-coupling for the vector mesons a
modification can be used:
\begin{equation}
\Gamma^\mu_{VNN}=g_{VN} \left( \gamma^\mu - q^\mu \frac{\gamma\cdot q}{q^2}
\right)
\quad.
\label{eq4_6_3}
\end{equation}
This vertex changes only the longitudinal part of the electromagnetic vertex
(thus contributes only to $F_3$) since
$j_\mu \Gamma^\mu_{VNN} = g_{VN} j_\mu \gamma^\mu$ and restores the WTI
since now $q_\mu \Gamma^\mu_{VNN} = 0$.
\section{Results}
\label{sec5}
Using dispersion techniques, H\"{o}hler and Pietarinen \cite{Hoehler} pointed
out the relation between pion-scattering phase shifts and electromagnetic
properties of nucleons. Because of a lack of higher lying resonances, it
cannot be expected that a simple loop expansion like the one developed here
can reproduce the required phase shifts for $\pi N$-scattering in the entire
energy regime under consideration. However, from the Lagrangians
(\ref{eq3_1_1}) and (\ref{eq3_1_2}) the phase shifts in the P11 and P33
channel can be successfully calculated close to threshold \cite{Oset}. On the
other hand, the $\pi \pi \to \pi \pi$-amplitude in the $J=T=1$-channel can be
best described by the $\rho$-resonance. All important low energy thresholds
are included for $W>m$ by the inelastic channels in the loop expansion, as
well as the thresholds for $q^2>4 m_\pi^2$ by using the correct momentum
dependent $\rho$-decay width as shown in fig.\ \ref{fig4_1}. So it is natural,
and no double counting is involved, to combine the loop expansion approach
(section \ref{sec3}) with the idea of vector meson dominance (section
\ref{sec4}).

The form factors are thus constructed in three steps. First, all diagrams of
fig.\ \ref{fig3_2} are calculated with the Lagrangians from section
\ref{sec3}. These diagrams form the contributions to $F^{s^\prime s}_{i,S}
(W^\prime,W;q^2)$ and $F^{s^\prime s}_{i,V} (W^\prime,W;q^2)$.
Next, these diagrams are calculated for a coupling to the neutral vector
mesons instead, using the respective experimentally determined coupling
constants for the $\rho^0$- and $\omega$-mesons to the hadrons and the
photons, resulting in the contributions to $F^{s^\prime s}_{i,\rho}
(W^\prime,W;q^2)$ and $F^{s^\prime s}_{i,\omega} (W^\prime,W;q^2)$.
Combined as in (\ref{eq4_3_2}), these contributions give the complete
irreducible vertex. Solving a system of linear equations (see appendix)
finally yields the irreducible form factors. The full form factors are
obtained by applying (\ref{eq2_5_1}).

The relevant coupling constants are determined by experiment and symmetry
considerations and are given in table \ref{tbl1}. The ratios between the
couplings of the pion to nucleons and deltas based on the
SU(2)$\times$SU(2)-considerations of \cite{WeiWirz} are
carried over to the $\rho$- and $\omega$-meson. The absolute values of the
$\rho NN$ and $\omega NN$ vertices are taken from \cite{Gross}. The
$\Delta N \gamma$ coupling constant $G_1$ is chosen to lie in between the two
values from \cite{Scadron}. So, besides $\Lambda$, no free paramter is
involved.
Because the cutoff only regularizes the divergent term the results still need
to be renormalized. In this paper the following renormalization scheme is
employed (quantities with superscript $R$ denote renormalized
quantities):
\begin{displaymath}
{W^*}^R(W) = W^*(W) + c_W ,\quad
{m^*}^R(W) = m^*(W) + c_m ,\quad
{f^{s^\prime s}_i}^R = Z f^{s^\prime s}_i \quad.
\end{displaymath}
The numbers $c_W$, $c_m$ and $Z$ are constants, independent of $q^2$ or $W$;
they can be determined from eq.\ (\ref{eq2_4_15}). One finds
\begin{equation}
{f^{++}_{1,p}}^R(m;q^2=0) =
\left. \frac{{W^*}^R-{m^*}^R}{W-m} \right|_{W=m}
,\quad \mbox{$p$ stands for proton} \quad.
\label{ren_1}
\end{equation}
For the renormalized theory one needs
\begin{eqnarray}
S^{-1}(p)S_0(p) &=& \Lambda^+(p) \frac{{W^*}^R-{m^*}^R}{W-m}
                + \Lambda^-(p) \frac{{W^*}^R+{m^*}^R}{W+m}  \nonumber \\
                &{{\displaystyle\longrightarrow}\atop{\scriptstyle W \to m}}& 1
                = \Lambda^+ + \Lambda^- \quad.
\label{eq5_ren}
\end{eqnarray}
Comparing the coefficients of the projection operators, it turns out, that in
this limit $\frac{{W^*}^R-{m^*}^R}{W-m}$ goes to 1, which is equivalent to the
statement that the full propagator has a pole with unit residue at $W=m$.
Therefore, one has ${W^*}^R(m) = {m^*}^R(m)$. It cannot be deduced, however,
that ${W^*}^R(m) = m$ and ${m^*}^R(m) = m$, because for $W \to m$ the
projector $\Lambda^-$ becomes zero itself, and thus there is no constraint on
${W^*}^R(m) + {m^*}^R(m)$.

For a pole of first order one has $Res(f(z);z_0) = \lim_{z \to z_0} (z-z_0)
f(z)$. With help of this, one can read off from (\ref{ren_1}) that
$1/{f^{++}_{1,p}}^R(m;0)$ is the value of the residue of the propagator and
must thus be equal to 1. This residue is adjusted with the wave function
renormalization constant $Z$. Therefore, one must choose
$Z=1/{f^{++}_{1,p}}(m;0)$. It turns out, that the full form factors are
renormalized automatically since, because of eq.\ (\ref{eq2_5_1}), the
wave function renormalization constant drops out. In the actual calculation
$Z = 0.37$ is found. For the self energies one gets $W^*(m) = m^*(m) = 0.9 m$,
thus $\Sigma(m) \approx 0.1 \mbox{GeV} (\gamma p - m)$, which is a reasonable
value.
%
%
\subsection{On-shell Form Factors}
Fig.\ \ref{fig5_1} shows the results for the full on-shell form factors
$F^{++}_{1,2}(q^2)$ for proton (a,c) and neutron (b,d). Note, that in this
case the full form factor is equal to the irreducible one. The squares are
calculated from the dipole fit to $G_E(q^2)$ and $G_M(q^2)$, which describe the
data for these momentum transfers up to a few percent in this energy range
\cite{dipfit}. In (b), in addition to the dipole fit, information about the
deuteron form factor has been taken into account. The two curves of squares
correspond to the extreme parametrizations of \cite{Becker} to indicate the
theoretical ambiguities in the description of the NN-potential for the
deuteron.

A severe test of the model, $F^{++}_{1,neutron}(q^2=0) = 0$, is well
fulfilled. For the proton, furthermore, $F^{++}_1$ is reproduced very well.
The radius of the proton comes out to be
\begin{displaymath}
\langle r^2 \rangle = 6 \left. \frac{\partial F^{++}_1(q^2)}{\partial q^2}
\right|_{q^2=0}
\quad \Rightarrow \quad
\langle r^2 \rangle^{1/2} = 0.81~\mbox{fm} \quad.
\end{displaymath}
$F^{++}_1$ for the neutron is comparable with the data; it falls in between
the theoretical uncertainties of the experimental analysis.
As discussed in the toy model of section \ref{sec4}, the VMD contribution to
the form factor accounts for a monopole shape only. Fig.\ \ref{fig5_1}
indicates, however, that in the full model a dipole shape is obtained.
Therefore, it is possible to conclude that in addition to the explicit
treatment of the vector meson pole terms, the $\pi$-loop corrections are thus
necessary to obtain the experimentally observed dipole shape of the form
factors. This conclusion has been discussed previously by Gari and
Kr\"{u}mpelmann \cite{GK} in a qualitative way, and is in agreement with the
findings of early dispersion theory treatment of the form factors. These
results give some confidence in the validity of the off-shell results.

Fig.\ \ref{fig5_1} c,d displays the behavior of $F^{++}_2(q^2)$. The shape of
the form factor is in agreement with the data in the momentum range plotted;
however, it does not show the correct behaviour at larger momenta.
Furthermore, the model cannot account for the full size of the anomalous
magnetic moment. The values found in this calculation are $\kappa_p=1.45$,
$\kappa_n=-1.65$, to be compared with the experimental values of
$\kappa^{exp}_p=1.79$, $\kappa^{exp}_n=-1.91$. As mentioned in section
\ref{sec4_2}, in the Lagrangian given in (\ref{eq4_1_2}) and (\ref{eq4_1_3}),
there is no contribution of the $\rho$-tensor coupling to $F_2$ for real
photons. This is in contrast to \cite{TT}, where, by multiplying the bare
vertex with $\kappa_\rho/2 \times m^2_V/(m^2_V - q^2)$, an anomalous magnetic
moment is induced. Such a procedure is highly questionable because it can be
argued that the tensor coupling of the $\rho$-meson to the nucleon is just due
to the loop corrections discussed here \cite{Flender}. To avoid any double
counting the tensor coupling must, therefore, not occur in the direct diagram.
In the present calculation, it has been taken into account only for the
diagrams with internal radiation to simulate the higher order corrections in a
schematic way. This is a crude approximation of the iteration scheme devised
in \cite{PFG}. The tensor coupling influences only the $q^2$-dependence of
$F_2$, not its magnitude at $q^2=0$. A number comparable to the magnetic
moment calculated here can be obtained by dividing the result of \cite{TT}
with meson cloud by their result for the bare vertex, giving roughly $2/3$.
Also \cite{NK} gets a number which is too small ($\kappa_p \approx 0.5$).

Fig.\ \ref{fig5_2} shows the irreducible form factors for projections to
negative energy states. $f^{+-}_1$ for the proton is not constrained to 1 by
(\ref{eq5_ren}). Its value of 0.9 at $q^2=0$ is directly related to the scalar
and vector self energies. Neither is $f^{+-}_1$ for the neutron constrained to
0. The magnetic form factors show a change of sign as one goes to higher
$q^2$. Their values at $q^2=0$ are drastically different from
$f^{++}_2(q^2=0)$. The full form factors $F^{+-}_i$ can be obtained by
rescaling $f^{+-}_i$ by 1/0.9. Even though these form factors never play a
role in experiments, they already indicate that all theoretical calculations
that rely on the asumption $F^{++}_i$=$F^{+-}_i$ are incorrect.

The models \cite{NK,TT,Bos} all approximately agree with each other for the
electric form factor. Since their magnetic moments are off by more than 50 \%
from experiment, an important process must have been missed in all these
calculations. There exists strong experimental evidence, that a large part of
the magnetic moment is due to spin flip transitions, especially to the
$\Delta$-resonance \cite{Burkert}. The present calculation shows, that about
half of the magnetic moment is carried by the diagram with the $\Delta \to N
\gamma$ decay mode, which is dominantly M1, a mode that was not included in
\cite{NK,TT,Bos}.

It must be stressed, that the underestimation of $\kappa$ does {\em not}
depend on the cutoff $\Lambda$ used in the calculation. Fig.\ \ref{fig5_3}
displays the $\Lambda$-dependence of $\kappa_p$ and $\kappa_n$. The magnetic
moments are optimal in the 1 GeV region, giving confidence in the combination
of the employed regularization and renormalization scheme. One clearly finds
for the larger values of $\Lambda$ that the magnetic moment decreases. Also for
small $\Lambda$ there is the onset of a decrease, which shows that the choice
of
the cutoff parameters cannot solve the disagreement. This is in strong
contrast to \cite{Flender}, where the tensor coupling increases monotonically
with $\Lambda$. The difference can be traced to the renormalization procedure
of
\cite{Flender}, where a subtractive renormalization is used, which is valid
only, if the wave function renormalization constant is close to 1; this
certainly is not the case in the present calculation ($Z = 0.37$). Note that
the agreement of state-of-the-art soliton models that include the
$\Delta$-resonance \cite{GCG} with the magnetic moment is of the same quality
as the agreement of the model presented here.

Bos et al.\ \cite{Bos} dress the nucleon with a scalar/isoscalar cloud. They
use a $\sigma$-meson with a mass of $m_\sigma = 0.8$ GeV. Since in this case
the photon only couples to the nucleon, this gives a large weight to the core
contributions. Still, the magnetic moment is only around 0.7. If a
contribution like this is included in the present calculation, there is some
effect, which is displayed by the dashed line in fig.\ \ref{fig5_3}. The
parameters chosen here are $m_\sigma = 0.56$ GeV, $g_\sigma = 11$. The effect
is due to a reduction of the wave function renormalization $Z$. This increases
the magnetic moments somewhat, however, without coming anywhere close to the
experimental point.
%
%
\subsection{Space Like half off-shell Form Factors}
Figs.\ \ref{fig5_4} and \ref{fig5_5} show the off-shell dependence of
$F^{+ \pm}_{1}(W;q^2)$ and $F^{+ \pm}_{2}(W;q^2)$ for the proton. In order to
see the
effects of the pure loop corrections all contributions due to VMD were
switched off in the upper panels. Part of the purpose of the following
discussion is to show, that not only in the time like electromagnetic form
factor one can see the influence of the $N\bar{N} \to \pi \pi$-scattering
\cite{Hoehler}, but also in the half off-shell form factors.

The left column of fig.\ \ref{fig5_4} shows $F^{++}_1$. For $W<m+2 m_\pi$ the
slope of the real part of $F^{++}_1$ increases with $W$, corresponding to an
increasing charge radius, which is in agreement with \cite{NK,TT}. Above
$m+2m_\pi$ the radius stays more or less
constant, the real part falls rather linearly with $-q^2$. The imaginary part
reflects the possibility that the incoming nucleon is far enough off-shell to
decay into final states of one or two pions plus a nucleon or a $\Delta$
during the electromagnetic scattering process. Because of the various
inelastic thresholds, the imaginary part is expected to depend on the phase
space of the decay products. This phase space dependence, however, is mixed
with self energy corrections due to the WTI, which themselves have imaginary
parts. Thus the imaginary part of $F^{++}_1$ does not show a clear behaviour
as $q^2$ and $W$ increase. The absolute value of the form factor including VMD
shows only little dependence on the off-shellness of the incoming proton for
$F^{++}_1$.

For $F^{+-}_1$ (right column of fig.\ \ref{fig5_4}) the situation is
different; above the $2\pi$-threshold it rises fast. Even after including VMD,
the changes due to off-shellness are clearly visible. Calculations of cross
sections for electromagnetic processes including off-shell nucleons and the
correct form factors must be performed in order to show the importance of the
negative energy components.

For the magnetic form factor $F^{+ \pm}_2$ (fig.\ \ref{fig5_5}) there is a
dramatic change in shape above the $2\pi$-threshold in the real part, whereas
the imaginary part mainly seems to grow with phase space. At $q^2=0$ the real
part of $F^{+ \pm}_2$ develops a maximum as a function of $W$ at the $2\pi$
threshold; the absolute value, on the other hand, increases rather smoothly.

A detailed analysis of the contributions of specific diagrams to the full form
factor reveals that the above effects are really caused by the $N^* \to \pi
\pi N$ channel, where $N^*$ denotes an excited nucleon. Fig.\ \ref{fig5_6}
shows the decomposition for $W=0.939$ GeV and $W=1.4$ GeV in comparison. The
full symbols show contributions of radiating pions (fig.\ \ref{fig3_2}c-e),
i.\ e.\ the meson cloud, the open symbols correspond to radiating baryons
(fig.\ \ref{fig3_2}b), i.\ e.\ the core. Squares describe diagrams with an
internal nucleon, circles represent diagrams with a propagating $\Delta$. The
solid line is the contribution of the direct diagram (fig.\ \ref{fig3_2}a) and
is counted as a core contribution. Contributions from the diagram of fig.\
\ref{fig3_2}f are indicated by dashed lines.

It is quite important to notice that for $F_1$ for both values of $W$ the cloud
contributions appear with different sign, leaving a more dominant core,
whereas for $F_2$ the core contributions, at least for $W=m$, tend to cancel.
So while $F_1$ is core dominated, $F_2$ is more influenced by the meson cloud.
The most important contribution to $F_2$ is the transition of the $\Delta$, in
agreement with the analysis of the Gerasimov-Drell-Hearn sum rule
\cite{Burkert}.

An obvious change in $F^{++}_1$, if one goes off-shell, is the increased
importance of the meson cloud from 15\% to 35\%. This is clearly an important
sign that the mesonic excitations must be carefully treated in all models for
off-shell form factors. Already in the on-shell case the contributions with a
propagating $\Delta$ are of some importance.

{}From the slopes one can deduce that the contributions of radiating pions
(cloud) reach further out than the ones of radiating baryons (core). Since the
different contributions show different slopes at $q^2=0$, which is especially
true for the off-shell case, it is not possible to give an adequate
parametrization of the $q^2$ dependence of the form factor using a single
parameter, if one wants to maintain a cloud/core picture to describe the
extension of the nucleon, like in the semiphenomenological models of
\cite{IJL,BRW}. This is even more clearly visible for the magnetic form
factor.

In going from the on-shell point to 1.4 GeV, the major changes come from the
diagrams \ref{fig3_2}c-e. For $F^{++}_1$ these fall off faster for 1.4 GeV;
for $F^{++}_2$ the shape is completely determined by these diagrams, all other
contributions stay rather constant. Since these diagrams are the only ones
that allow for the $2\pi$-decay, it must be concluded that the $N^* \to \pi\pi
N$ channel is responsible for these effects rather than the $N^* \to \pi
\Delta$ channel which is already open, too. This also indicates that the
effects described above are not spurious and will remain if the $\Delta$ is
treated in a more realistic way by allowing for a finite decay width
(cf. section \ref{sec3_1}).

%
%
\subsection{Time Like Form Factors}
Figs.\ \ref{fig5_7} and \ref{fig5_8} show the electric and magnetic form
factors for the proton, respectively, in the entire range covered by this
calculation. The contour lines show $\log |F_{1,2}^{++}(W^2;q^2)|$. Various
thresholds are indicated by dotted lines. The region, which is not accessible
by experiments ('unphysical region') is indicated by the gray area; all
dispersion relation approaches rely on analytical continuation into this
region.

It must be emphasized, that at this level all inelasticities up to the $2\pi$
channel are included in the vertex either by the $\rho \to \pi\pi$ decay, or
by $N^* \to X + n\pi$ $(n=1,2)$. While $F_1$ shows almost no changes as one
goes off shell, for $F_2$ the influence of inelastic channels remains visible.
Especially the onset of $2\pi$ production influences the form factor.

It is due to the WTI that $F_1$ does not change very much when going
off-shell. Each off-shell effect in the vertex is nearly canceled by the self
energy corrections (\ref{eq2_5_1}), which have the same thresholds.
Interestingly, this cancellation seems to be independent of $q^2$. This is not
entirely true (see last subsection), however, the $q^2$ dependence is so much
dominated by VMD that other effects, which depend on $q^2$, are hardly
visible. This is a very important result. If VMD exists for nucleons as well
as for pions, it will clearly be observable, since it is not reduced for
off-shell nucleons in the case of $F_1$, and is still dominant for $F_2$.

For $W=2$ GeV fig.\ \ref{fig5_sv} displays the scalar and vector contributions
of the direct coupling graphs and of vector meson graphs to the form factors.
{}From this information
the proton or neutron form factor can be constructed. For $F_{1 S/V}$ it
shows a smooth behaviour; the minimum in the vector imaginary part is due to
the $\pi^+ \pi^-$-channel. The pure photon channels are clearly least
important as compared to the vector meson channels. The figure shows the broad
rho- and the narrow omega-resonance. Since both mesons couple to the nucleon
with about equal strength, the $\omega$ contribution is an order of magnitude
larger than the $\rho$ contribution at peak level because of the $\omega$'s
small width. Note that in the case of $F_2$ for the vector meson contributions
the role of imaginary and real part has changed, indicating that the imaginary
part of the loop expansion without VMD is significantly larger than its real
part. This is consistent with the situation in the space like sector (fig.\
\ref{fig5_5}) and gives rise to interference. For $F_2$ the contribution of
the $\rho$-meson if of larger importance than for $F_1$.

To show once again that VMD can be observed even in an experiment involving
a half off-shell vertex, a cut of fig.\ \ref{fig5_7} is plotted in fig.\
\ref{fig5_9}. The absolute magnitude of the single contributions in the
photon-, $\omega$- and $\rho$-channels as well as the absolute magnitude of
the coherent sum is displayed for time like momentum transfers for the proton
and the neutron at $W=2$ GeV. Despite the small width of the
$\omega$-resonance it is not possible to resolve the $\rho$-contribution in
$F_1$. The sum is almost exclusively exhausted by the $\omega$-contribution,
while the $\rho$ on the other hand is only visible as a broad background. On
the other hand, for $F_2$ the $\rho$-component is not much smaller than the
$\omega$-component, therefore, subtle interference effects show up between the
$\rho$- and $\omega$-channel, which might be accessible in experiments on both
protons and neutrons.

Experiments on dilepton production in $p+A$ reactions, where the $\pi^+
\pi^-$-annihilation channel is surpressed, may thus offer a chance to study
medium modifications of the $\omega$-meson from the electromagnetic properties
of the nucleon in the same way as medium modifications of the $\rho$-meson
from electromagnetic properties of the pion. Because the nucleon's form
factors are larger than the one of the pion (at peak position $|F_\pi| \approx
7$), it is visible even in heavy ion collisions above the background from
$\pi^+\pi^-$ annihilation \cite{Wolfpriv}; however, a very good resolution of
the experimental apparatus is required to resolve it.

%
%
\section{Summary}
\label{sec6}
To study the production of dileptons in high energy nuclear reactions one
needs information about the half off-shell time like electromagnetic form
factors, which are until now unknown. We have, therefore, constructed a
dynamical model based on a hadronic framework to calculate the electromagnetic
form factors for momentum transfers of the photon of $-1$ GeV$^2 < q^2 < 1$
GeV$^2$ and for nucleons with $-1$ GeV$^2 < p^2 < 4$ GeV$^2$.

Starting from an expansion of the nucleon propagator in pion loops, the
electromagnetic vertex is constructed by inserting external photons to each
charged particle line, thus obeying the constraints due to the WTI. Together
with the concept of VMD, this approach includes at least schematically all
findings of the spectral analysis of the late 1960 \cite{Hoehler}. It is
possible to maintain a cloud/core picture that proved successful in
semiphenomenological models \cite{IJL,BRW}. It is important to include the
$\Delta$-resonance to reproduce the magnetic properties, which is in accord
with an analysis of the Gerasimov-Drell-Hearn sum rule \cite{Burkert}. The
coupling constants and the cutoff are chosen in agreement with meson exchange
models and symmetry considerations \cite{WeiWirz}.

The momentum dependence of the space like on-shell form factors is reproduced.
The charge radius of the proton is found to be $r=0.81$ fm. The electric form
factor of the neutron falls in between the uncertainties of the data analysis.
The magnetic moments are $\kappa_p=1.45$ and $\kappa_n=-1.65$, which is much
closer to the experimental value than in comparable calculations \cite{NK,TT},
but still too small. The better agreement can be traced to the $\Delta \to
N\gamma$ decay process, which occurs in neither of the above cited works. A
decomposition of the form factors shows that for $F_1$ the core contributions
dominate, while for $F_2$ the cloud is more important. It also shows that
inelastic thresholds influence the form factors. However, for $F_1$ the
dependence is only weak because it is compensated by self energy corrections
required by gauge invariance. For $F_2$ threshold effects remains visible.

The weak off-shell effects enable one to study VMD in the experimentally
accessible region of the $(W,q^2)$ plane. It turns out, that only the signal
of the $\omega$-meson can clearly be extracted, the $\rho$-meson contributions
only result in a broad background. As an effect one has the possibility to
study medium-effects on the $\omega$ very clearly by measuring the
electromagnetic form factor in $p+A$- and even in heavy ion collisions.

It is found, that calculations which are based on the assumption that
$F^{++}_i = F^{+-}_i$ must be rechecked.

The model for the vertex used here certainly leaves room for improvements.
Taking
into account higher resonances, especially of spin 3/2, as well as heavier
mesons will help to improve the magnetic moments. The finite decay width of
the $\Delta$ also needs to be incorporated. But since the most prominent
threshold effect is due to $N^*\to N \pi\pi$, it is expected, that the
off-shell behaviour of the form factors will not change appreciably.

\subsection*{Acknowledgments}

Interesting discussions with members of the DLS group and with Gy\"{o}rgy Wolf
are
greatly appreciated. A hint of T.\ Maruyama, concerning numerical integration,
was extremely helpful. This work was supported BMFT and GSI Darmstadt.

\newpage

\begin{appendix}
\section{Numerical Details}

The sum of all Feynman diagrams in fig.\ \ref{fig3_2} results in the
irreducible vertex. To obtain the form factors from the most general vertex
(\ref{eq2_2_3}) the projection on positive/negative energy must be performed
for in- and outgoing nucleons:
\begin{equation}
\Lambda^{s^\prime}(p^\prime) \Gamma^\mu (p^\prime,p) \Lambda^s(p)
= \Lambda^{s^\prime}(p^\prime)
\sum_{i=1}^3 f^{s^\prime s}_{i} (W^\prime,W;q^2) {\cal O}^\mu_i
\Lambda^s(p) \quad.
\label{eq_a_1}
\end{equation}
Performing traces over contractions of the vertex with three linear
independent 4-vectors $v_1^\mu = P^\mu = {p^\prime}^\mu + p^\mu$,
$v_2^\mu = q^\mu = {p^\prime}^\mu - p^\mu$, $v_3^\mu = \gamma^\mu$ gives
\begin{equation}
\tilde{T}_i
    = \mbox{Tr } v^\mu_i \Lambda^{s^\prime}(p^\prime) \Gamma_\mu \Lambda^s(p)
    = \sum_{j=1}^3 a_{ij} F^{s^\prime s}_j \quad.
\label{eq_a_2a}
\end{equation}

The same procedure is performed with the sum $S_\mu$ of all contributing
Feynman diagrams:
\begin{equation}
T_i = \mbox{Tr } v^\mu_i \Lambda^{s^\prime}(p^\prime) S_\mu \Lambda^s(p)
\quad.
\label{eq_a_2b}
\end{equation}
To obtain the form factors one must equate (\ref{eq_a_2a}) and (\ref{eq_a_2b}):
\begin{equation}
T_i = \tilde{T}_i = \sum_{j=1}^3 a_{ij} F^{s^\prime s}_j \quad.
\label{eq_a_2c}
\end{equation}

While the expressions $\tilde{T}_i$ can be obtained analytically, the
quantities $T_i$ must be calculated numerically. They can be decomposed into
integrals of the following form:
\begin{displaymath}
T_i = \sum_{d} \int \frac{d^4k}{(2\pi)^4} \left[\sum_{u,v,w}
c^{d s^\prime s}_{u,v,w} ({p^\prime}^2,p^2,k^2)
(p^\prime k)^u (pk)^v (k^2)^w / N_d(p^\prime,p,k) \right] \quad.
\end{displaymath}
The index $d$ labels the different diagrams contributing to $S_\mu$.
$N_d(p^\prime,p,k)$ contains the denominator of the propagators in diagram
$d$.

The coefficients $c^{d s^\prime s}_{u,v,w} ({p^\prime}^2,p^2,q^2)$ as well as
the $a_{ij}$ in eq.\ (\ref{eq_a_2a}) are
calculated using the high energy package of REDUCE \cite{reduce}.

All integrals are of the form
\begin{displaymath}
I(p^\prime,p) = \int \frac{d^4k}{(2\pi)^4}
  \frac{(p^\prime k)^u (p k)^v (k^2)^w}
       {\prod\limits_{p=1}^N \left((a_p-k)^2-m^2_p+\mbox{i} \epsilon \right)}
       \quad,
\end{displaymath}
where $a_p$ and $m_p$ are the specific momenta and masses of the propagators
in the diagrams. The number $N$ of factors in the denominator depends on the
choice of monopole or dipole cutoff and ranges from 4 to 7. The integrals are
performed numerically.

$I(p^\prime,p)$ is solved in the restframe of the outgoing nucleon, which
always exists in case of half off-shell kinematics. The integration of $k^0$
is done by contour integration around the poles of
$\prod_{p=1}^N ((a_p-k)^2-m^2_p+\mbox{i} \epsilon)^{-1}$. Let $k^0_i$ denote
the poles in the upper half plane. Since not all poles are of first order, for
each $\vec{k}$ the residue must be calculated by taking numerical derivatives:
\begin{equation}
I(p^\prime,p) = \mbox{i} \int \frac{d^3k}{(2\pi)^3} \sum_{i=1}^n Res(k^0_i)
\quad,
\label{eq_a_3}
\end{equation}
\begin{displaymath}
Res(k^0_i) = \frac{1}{(n_i-1)!} \frac{d^{n_i-1}}{{dk^0}^{n_i-1}}
\left[ \frac{(p^\prime k)^u (p k)^v (k^2)^w (k^0-k^0_i)^{n_i}}
{\prod\limits_{p=1}^N \left( (a_p-k)^2-m^2_p+\mbox{i} \epsilon \right)}
\right]_{k^0=k^0_i} \quad,
\end{displaymath}
where $n_i$ is the order of the pole at $k^0_i$. For space like momentum
transfer the sum of the residues is a well behaved function of $\vec{k}$.
However, in the time like region it still has poles because of the physical
inelasticities for $q^2 \ge 4 m_\pi^2$ and $p^2 \ge (m+m_\pi)^2$. These poles
are treated with a subtraction technique, which will be described below. Since
the momenta $p^\prime$ and $p$ can be chosen such that $I$ is invariant
under rotations about the $z$-axis, the $\phi$ integration is trivial.
The remaining integration is two dimensional. From now on $k$ denotes
$|\vec{k}|$.
\begin{displaymath}
I(p^\prime,p) = \int\limits_{-1}^1 dx \int\limits_0^\infty dk
\frac{f(k,x)}{g(k,x)} \quad \quad\mbox{with } \quad x=\cos \theta
\end{displaymath}
Let $k_i(x)$ be defined by $g(k_i(x),x)=0$ $(i=1,2)$. The $k_i(x)$ are complex
functions of $x$. For $x \ge x_0$ the imaginary part becomes $\pm \epsilon$.
Below $x_0$ an ordinary integration can be used. Above $x_0$ the integral
$I(p^\prime,p)$ splits into a principle value integral and an imaginary part.
\begin{eqnarray}
\int\limits_0^\infty dk \frac{f(k,x)}{g(k,x)} &=&
\int\limits_0^\infty dk \frac{f(k,x)}{g_R(k,x)} \frac{1}
{\prod_{i=1}^2 [k_i(x,\epsilon=0)-k+\mbox{i} \epsilon Re(dk_i/d\epsilon)]}
                                                              \nonumber \\
&=& P \int\limits_0^\infty dk \frac{f(k,x)}{g_R(k,x)}
      \frac{1} {[k_1(x,\epsilon=0)-k][k_2(x,\epsilon=0)-k]}   \nonumber \\
&-& \mbox{i} \pi \frac{f(k_1(x),x)}{g_R(k_1(x),x)}
             \frac{dk_1/d\epsilon}{k_2(x)-k_1(x)}
             \Theta(k_1(x))                                   \nonumber \\
&+& \mbox{i} \pi \frac{f(k_2(x),x)}{g_R(k_2(x),x)}
             \frac{dk_2/d\epsilon}{k_1(x)-k_2(x)}
             \Theta(k_2(x))
\label{eq_a_4}
\end{eqnarray}
Since $P \int_0^\infty 1/(k^2-k_0^2) dk = 0$, the principle value integral is
treated as follows:
\begin{eqnarray*}
\lefteqn{P\int\limits_0^\infty dk \frac{f(k,x)}{g(k,x)} = } && \\
&& = \int\limits_0^\infty dk \left[
 \frac{f(k,x)}{g(k,x)}
-\sum_{i=1}^2 \frac{f(k_i(x),x)}{k^2-k_i^2(x)}
       \left. \frac{k^2-k_i^2(x)}{g(k,x)} \right|_{k=k_i(x)} \Theta(k_i(x))
\right] \quad.
\end{eqnarray*}
The integrand is now finite for all $k$.

If $p^2 \ge (m+2 m_\pi)^2$, then in some diagrams both functions $k_i(x)$
contribute to the imaginary part. In this case there is a remaining
singularity of the type $1/\sqrt{1-(x/x_0)^2}$ at $x=x_0$. The integral is
thus converging, but numerically unstable. It can be treated by a trick
similar to the one used above:
\begin{displaymath}
\int\limits_{-|x_0|}^{|x_0|} \frac{\varphi(x)}{\sqrt{x_0^2-x^2}} dx
= \int\limits_{-|x_0|}^{|x_0|}
  \frac{\varphi(x)-\varphi(x_0)}{\sqrt{x_0^2-x^2}} dx
  + \pi \varphi(x_0)
\end{displaymath}

After having calculated all integrals in $T_i$, the equations (\ref{eq_a_2c})
are solved for the invariant form factors.

Numerical inaccuracies can occur at various points. For example, in
(\ref{eq_a_3}) only the sum of all residues falls off fast enough in $k$ so
that the integral converges. There is a delicate cancellation of the summands
which can be shown analytically that is hard to reproduce by taking numerical
derivatives. Furthermore, for large $k$ the poles come near to each other,
sometimes too close to take the numerical derivative with good enough
accuracy. This is worked around by not evaluating the integrand at too large
values of $k$. A cutoff is chosen dynamically if the integrand is small
enough. For higher values of $k$ asymptotic behaviour is assumed and the
integral is solved analytically. Unfortunately under this assumption the WTI
suffer. They are only fulfilled at the 1\% level.

In (\ref{eq_a_4}) problems arise when $|x_0|$ is close to 1. Then too few
gridpoints of the integration mesh contribute to the imaginary part. In these
cases additional grid points are created and integration weights are
redistributed such that the integral is performed over at least 10 grid points
in $\cos \theta$-direction.

The numerical difficulties are best under control for a very simple
integration with equally spaced grid points of equal weights. To obtain an
acceptable accuracy up to 400 points in radial direction and 100 points in
$\cos \theta$ direction are used.

\newpage
\section{Figure Captions}
\newcounter{figno}
\begin{list}{\underline{Fig.\arabic{figno}}:}{\usecounter{figno}\setlength{\rightmargin}{\leftmargin}}
\item
Expansion of the full nucleon propagator in first order in pion
lines. The nucleon propagator is displayed by the solid single line, the
$\Delta$-propagator by the solid double line, the full propagator by the solid
line with the fat dot, the pion propagator by the dashed line.
\label{fig3_1}

\item
Loopwise expansion of the irreducible nucleon vertex. (a) is the free
vertex, (b) and (c) come from coupling of photons to charged hadron lines
of the propagator in fig.\ \protect\ref{fig3_1}, (d) and (e) arise from
contact terms due to pseudo-vector coupling of pions to nucleon and $\Delta$,
(f) is the contribution of the decay $\Delta \to N \gamma$.
\label{fig3_2}

\item
Additional diagrams to satisfy WTI if a cutoff is introduced in
the $\pi N$- or the $\pi \Delta$-vertex. The double dashed line represents
the 'propagator of the cop'.
\label{fig3_3}

\item
Possible cuts of the loop diagrams. The cut in (a) corresponds to
the decay of the off-shell internal nucleon to an on-shell nucleon and a
pion, the cut in (b) corresponds to $\pi^+ \pi^-$-annihilation if the photon
has high enough invariant mass, or to two pion production if the incoming
nucleon is far enough off-shell.
\label{fig3_4}

\item
Electromagnetic form factor of the pion. The data are taken from
\protect\cite{Quenzer,pifspace,piftime1}, the solid line is obtained using
(\protect\ref{eq4_2_1}) and adding a similar contribution for the $\Phi$-
meson.
\label{fig4_1}

\item
$F^{++}_{1,2}(m,m;q^2)$ for (a,c) proton and (b,d) neutron. The solid
line is the model calculation, the symbols represent experimental results as
explained in the text. For $F_2$ the experiment is rescaled to the anomalous
magnetic moment of this calculation.
\label{fig5_1}

\item
$f^{+-}_{1,2}(m,m;q^2)$ for (a,c) proton and (b,d) neutron.
\label{fig5_2}

\item
The anomalous magnetic moment for proton and neutron as function of
the cutoff parameter $\Lambda$. The solid and dashed lines display the result
with and withot contributions of the $\sigma$-meson respectively.
\label{fig5_3}

\item
Half off-shell form factors of the proton in the space like region.
The left panel displays $F^{++}_{1}(W;q^2)$ for several incident invariant
masses.
The curves with symbols are below the $2 \pi$-threshold, the curves without
symbols are above. The two upper plots show real and imaginary part of the
form factor without VMD contributions, while the lower plot shows the absolute
value including VMD. The right panel displays $F^{-+}_{1}(W;q^2)$ in the same
way.
\label{fig5_4}

\item
Same as fig.\ \protect\ref{fig5_4} but for $F^{\pm +}_{2}(W;q^2)$.
\label{fig5_5}

\item
Detailed analysis of the contribution of the single diagrams in the
$W=1.4$ GeV case (right panel) compared to the on-shell case (left panel) for
the proton. Lines with full symbols correspond to meson cloud contributions,
lines with open symbols and the full line are core contributions. The dashed
line represents the $\Delta \to N \gamma$ decay contribution.
\label{fig5_6}

\item
$F^{++}_1$ for the proton in the full range of applicability of the
model. The contour lines are steps of 0.1 in $log_{10} | F |$. Various
thresholds are indicated by dashed lines. The experimentally unaccessible
region is marked as the grey area.
\label{fig5_7}

\item
Same as fig.\ \protect\ref{fig5_7}, but for $F^{++}_2$.
\label{fig5_8}

\item
Real and imaginary part of scalar and vector contributions to the
electric form factor for $W=2$ GeV. The full lines display the isoscalar
contributions, the dashed lines display the isovector ones. Symbols indicate
imaginary parts. The upper plots show the scalar and vector contributions of
the direct photon coupling. The plots below show contributions of $\omega$-
and $\rho$-mesons seperately.
\label{fig5_sv}

\item
Decomposition of $|F^{++}_{1,2}|$ for the proton (left) and neutron
(right) at $W=2$ GeV into components coming from the direct photon vertex, the
$\rho$-meson vertex and the $\omega$-meson vertex. The sum is the result of a
coherent superposition of the single contributions.
\label{fig5_9}

\end{list}
\end{appendix}
\newpage

\begin{table}[h]
\begin{center}
\begin{tabular}{|l|c|c|l|}
 & $g_{\omega N}/g_{\omega \gamma}$ & $g_{\rho N}/g_{\rho \gamma}$ & \\[3mm]
\hline \hline
H\"{o}hler et al. \cite{Hoehler2} & 0.52\phantom{0} & 0.98\phantom{0} &
includes
$\pi N$ phase shifts in $\rho$
already \\
\hline
Dubni\v{c}ka \cite{Dub} & 0.376 & 0.418 & remaining strength in higher poles
\\
\hline
Gari et al. \cite{GK} & 0.377 & 0.411 & remaining strength in direct coupling
\end{tabular}
\end{center}
\caption{Comparison of coupling constants in VMD like polefits}
\label{tbl2}
\end{table}

\begin{table}[h]
\begin{displaymath}
\begin{array}{c}
\begin{array}{lcrllcrllcr}
g_{NN\pi} &=& 13.45 &,& g_{N\Delta\pi} &=& 22.86 &,& g_{\Delta\Delta\pi} &=&
10.76\\
g_{NN\omega} &=&  8.34 &,& g_{N\Delta\omega} &=& 14.18 &,&
g_{\Delta\Delta\omega} &=&  6.67\\
g_{NN\rho} &=&  7.26 &,& g_{N\Delta\rho} &=& 12.35 &,& g_{\Delta\Delta\rho} &=&
5.81
\end{array}
\\
\begin{array}{c}
G_1 = 2.5~\mbox{GeV}^{-1}\\
\kappa_\rho = -6 \quad,\quad \kappa_\omega=0\\
\Lambda = 1.2~\mbox{GeV}
\end{array}
\end{array}
\end{displaymath}
\caption{Coupling constants}
\label{tbl1}
\end{table}

\end{document}